\documentclass[%
superscriptaddress,
twocolumn, 
showpacs,
 amsmath,amssymb,
floatfix,
]{revtex4-1}
\usepackage[utf8]{inputenc}
\usepackage{graphicx,color,soul} 
\usepackage{amsmath}
\usepackage{hyperref}
\usepackage{upgreek}

\begin{document}

\newcommand{\ms}[1]{\mbox{\scriptsize #1}}
\newcommand{\msb}[1]{\mbox{\scriptsize $\mathbf{#1}$}}
\newcommand{\msi}[1]{\mbox{\scriptsize\textit{#1}}}
\newcommand{\nn}{\nonumber} 
\newcommand{\dg}{^\dagger}
\newcommand{\smallfrac}[2]{\mbox{$\frac{#1}{#2}$}}
\newcommand{\ket}[1]{| {#1} \rangle}
\newcommand{\bra}[1]{\langle {#1} |}
\newcommand{\out}[1]{| {#1} \rangle \langle {#1} |}
\newcommand{\kr}{|r\rangle}
\newcommand{\br}{\langle r|}
\newcommand{\ba}{\langle\alpha |} 
\newcommand{\bb}{\langle\beta |} 
\newcommand{\ka}{|\alpha\rangle}
\newcommand{\kb}{|\beta\rangle}
\newcommand{\pfpx}[2]{\frac{\partial #1}{\partial #2}}
\newcommand{\dfdx}[2]{\frac{d #1}{d #2}}
\newcommand{\half}{\smallfrac{1}{2}}
\newcommand{\s}{{\mathcal S}}
\newcommand{\jord}{\color{red}}
\newcommand{\kurt}{\color{blue}}
\newtheorem{theo}{Theorem} \newtheorem{lemma}{Lemma}

\title{Steady-state microwave mode cooling with a diamond NV ensemble}

\author{Donald P. Fahey} 
\affiliation{DEVCOM Army Research Laboratory, Sensors and Electron Devices Directorate, Adelphi, Maryland 20783, USA}

\author{Kurt Jacobs}
\affiliation{DEVCOM Army Research Laboratory, Sensors and Electron Devices Directorate, Adelphi, Maryland 20783, USA}
\affiliation{Department of Physics, University of Massachusetts at Boston, Boston, MA 02125, USA}

\author{Matthew J. Turner}
\affiliation{Quantum Technology Center, University of Maryland, College Park, Maryland 20742, USA}
\affiliation{Department of Electrical and Computer Engineering, University of Maryland, College Park, Maryland 20742, USA}

\author{Hyeongrak Choi}
\affiliation{Department of Electrical Engineering and Computer Science, Massachusetts Institute of Technology, 77 Massachusetts Avenue, Cambridge, Massachusetts 02139, USA}

\author{Jonathan E. Hoffman} 
\affiliation{DEVCOM Army Research Laboratory, Sensors and Electron Devices Directorate, Adelphi, Maryland 20783, USA}

\author{Dirk Englund} 
\affiliation{Department of Electrical Engineering and Computer Science, Massachusetts Institute of Technology, 77 Massachusetts Avenue, Cambridge, Massachusetts 02139, USA}

\author{Matthew E. Trusheim} 
\affiliation{DEVCOM Army Research Laboratory, Sensors and Electron Devices Directorate, Adelphi, Maryland 20783, USA} 
\affiliation{Department of Electrical Engineering and Computer Science, Massachusetts Institute of Technology, 77 Massachusetts Avenue, Cambridge, Massachusetts 02139, USA}

\begin{abstract}
A fundamental result of quantum mechanics is that the fluctuations of a bosonic field are given by its temperature $T$~\cite{bose1924plancks}. An electromagnetic mode with frequency $\omega$ in the microwave band has a significant thermal photon occupation at room temperature according to the Bose-Einstein distribution $\bar{n} = k_BT / \hbar\omega$. The room temperature thermal state of a 3 GHz mode, for example, is characterized by a mean photon number $\bar{n} \sim 2000$ and variance $\Delta n^2 \approx \bar{n}^2$. This thermal variance sets the measurement noise floor in applications ranging from wireless communications to positioning, navigation, and timing to magnetic resonance imaging~\cite{Cakaj2011satellite,Ho2007LinkAO}. We overcome this barrier in continuously cooling a ${\sim} 3$ GHz cavity mode by coupling it to an ensemble of optically spin-polarized nitrogen-vacancy (NV) centers in a room-temperature diamond. The NV spins are pumped into a low entropy state via a green laser and act as a heat sink to the microwave mode through their collective interaction with microwave photons. Using a simple detection circuit we report a peak noise reduction of $-2.3 \pm 0.1 \, \textrm{dB}$ and minimum cavity mode temperature of $150 \pm 5 \textrm{K}$. We present also a linearized model to identify the important features of the cooling, and demonstrate its validity through magnetically tuned, spectrally resolved measurements. The realization of efficient mode cooling at ambient temperature opens the door to applications in precision measurement and communication, with the potential to scale towards fundamental quantum limits.
\end{abstract}

\maketitle

\section{Main} 

Traditional methods of thermal noise reduction in the microwave regime have relied on cooling the entire apparatus~\cite{krinner2019cryo,putz2014protection,krantz2019superconducting,Montazeri2016cryoamp}. A more direct route is to cool the electromagnetic modes themselves. This is possible when a single mode in a cavity has a very low heat capacity and long thermalization time compared with the constituent material. Targeted cooling of a subsystem has long been employed in atomic gases~\cite{lett1988lasercooling,regal2012tweezercooling}, trapped ions~\cite{king1998coolions,rohde2001coolions}, and micromechanical oscillators~\cite{metzger2004microlever,schliesser2006backaction}, but only recently has the selective cooling of an electromagnetic mode of interest been explored. In Wu et al.~\cite{OxborrowPentacene}, the effectively low temperature of a polarized spin bath of pentacene was used to remove the thermal energy from a cavity mode through their mutual coupling, but was limited to pulsed operation. In this work, we experimentally realize steady-state microwave cavity mode cooling and explain its mechanism with an analytic theory. Our quantum mechanical treatment of this ``spin refrigerator'' indicates an optimal parameter regime that depends both on cavity coupling parameters and spin ensemble inhomogeneity. 

\begin{figure}[t]
\centering
\includegraphics[width=0.5\textwidth]{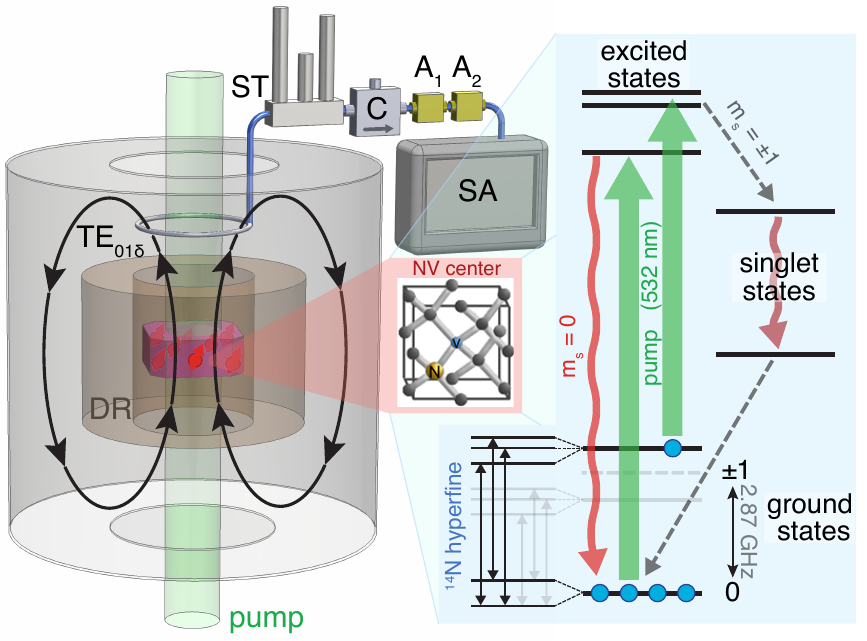}
\caption{Cooling a microwave mode with a diamond NV- ensemble: a purple diamond inside a dielectric resonator (DR) at room temperature is pumped with a green laser. The hyperpolarized NV- ground state spins are coupled to the $\textrm{TE}_{01 \delta}$ cavity mode by shifting them into resonance with magnetic bias coils (not shown). The resulting reduction in mode temperature is probed by a coupling loop and spectrum analyzer (SA) via a stub tuner (ST), circulator (C), and two stage amplifier ($\textrm{A}_{1}$ and $\textrm{A}_{2}$).}
\label{fig:first}
\end{figure}

We consider the system presented schematically in Fig.~\ref{fig:first}. Here a microwave mode is confined within a dielectric resonator, where it couples to the NV spin ensemble in a co-located diamond via magnetic dipole interaction $H_\text{int} = -\hat{\mu} \cdot \hat{B} $, where $\hat{\mu}$ is the transition (magnetic) dipole moment and $\hat{B}$ is the magnetic field operator of the cavity mode \cite{choi2021ultrastrong}. The negatively charged NV- center possesses a triplet ground state with spin transition $m_s =0 \leftrightarrow m_s = \pm 1$ at $2.87$ GHz and 290 K. An external magnetic bias field breaks the degeneracy of the $m_s = \pm1$ NV ground state spin levels and tunes one of the spin transitions near resonance with the microwave mode. Simultaneously, each NV center undergoes spin polarization under off-resonant laser illumination via the excited orbital state and subsequent intersystem crossing and relaxation \cite{doherty2013nitrogen}. This process, in competition with direct spin relaxation, produces a net steady-state polarization, equivalent to a low temperature NV spin population. The coupling to the microwave resonator transfers thermal excitations in the cavity mode to the cold spin bath. 

For a polarized spin ensemble in the weak-excitation limit, where the mean intracavity photon number $\bar{n}$ normalized by the number of cavity-coupled NV centers ($N$) is small, $\bar{n}/N\sim 0$, we can approximate the spins as harmonic oscillators~\cite{yamamoto1999mesoscopic}. This `linearized' approach, also called the Dicke or Tavis-Cummings model \cite{hepp1973superradiant}, has been widely used for describing the coupling between spin ensembles and the cavity mode. Cavity mode cooling in this approximation can be understood following the quantum Langevin equation~\cite{Gardiner85,Jacobs14}. The interaction of a system with a bath of harmonic oscillators (spins) that has a Lorentzian distribution of frequencies can be described exactly by an interaction with a single spin pseudo-mode that is itself damped by coupling to the standard Markovian thermal bath of oscillators~\cite{Imamoglu94, Garraway97}. The rate at which this single pseudo-mode is damped to the Markovian bath is given by the frequency spread of the original spin bath. Since the decoherence rate of the NVs and the damping rate (polarization rate) per NV center provided by the pump laser are small compared to the inhomogeneous broadening, it is the latter that provides the irreversible damping of the spin system to the cold bath. 

We parametrize the system in a linear model following the prescription above: using the collective interaction rate between the cavity and the spins, $g$, the internal loss rate and the output coupling rate of the cavity mode, denoted respectively by $\gamma$ and $\kappa$ (and through which the cavity mode is heated by a room temperature bath), the inhomogeneous broadening of the NV ensemble, $r$, and the polarization $P = |P_{\ms{0}}-P_{\ms{1}}|$ of the spin bath provided by the continual laser pumping, where $P_{0}$($P_{1}$) is the population of $m_s=0(1)$ normalized by the total number of NVs in either $m_s=0$ and $m_s=1$ states. We give the main results of our linear model here, while the full theoretical analysis is given in Supplementary Information.

The mean intracavity photon number, $\bar{n}$, evaluates to a weighted sum of the room-temperature photon number $n_T = [\exp{\frac{\hbar\omega}{k_BT}}-1]^{-1}$ and the effective `cold spin bath' photon number, $n_{\ms{c}}=\frac{1}{2}\left(\frac{1 - P}{P}\right)$: 
\begin{equation}
        \bar{n}  = \cos^2(\theta)  n_T + \sin^2(\theta)  n_{\ms{c}}     ,
\end{equation}
where $\sin^2(\theta)\equiv \xi r /(\kappa +  \gamma  + \xi r )$ denotes the `cooling ratio' under an effective cooling rate $\xi r$. This rate is the inhomogeneous broadening of the spins reduced by the dimensionless parameter $\xi = (1 + \left[\frac{r(r + \kappa + \gamma)}{4g^2} \right])^{-1}$. Highly effective cooling is obtained when $r \gg \kappa + \gamma$ and $g \gtrsim r$. In particular the theory reveals that it is the inhomogeneous broadening of the ensemble, along with the ensemble coupling $g$, that primarily governs the rate at which heat is extracted from the ensemble. The per-spin polarization rate $\Gamma_p$ induced by the laser only determines the temperature of the spin bath rather than the rate of extraction of thermal power, and as such can be much smaller than $r$ and $g$.

The electromagnetic mode temperature is obtained through the noise power spectrum. The reduction in temperature depends on the measured electromagnetic mode frequency as well as the spin-cavity detuning. The linear theory gives the cavity output power spectrum as
\begin{widetext}
\begin{align}\label{eq:result}
   N_{P}(\omega) 
    & = n_T + (n_c - n_T) \left[\frac{ \kappa r g^2     }{ \left(\frac{r^2}{4} + \omega^2 \right) \left( \frac{(\kappa + \gamma)^2}{4} +(\omega-\Delta)^2 \right) + g^2  \left(\frac{r(\kappa + \gamma)}{2} -2\omega(\omega-\Delta) \right) + g^4  }   \right] , 
\end{align}
\end{widetext}
where $\omega = 0$ is the cavity frequency and $\Delta$ gives the cavity-spin detuning. When the spins are resonant the frequency at which the noise reduction is maximal occurs at the cavity frequency unless $8 g^2 > r^2 + (\kappa +\gamma)^2$ in which case the maximum happens for $\omega^2 = g^2 - [r^2 + (\kappa +\gamma)^2]/8$. In the former case the maximum noise reduction is $2 \kappa r g^2 (n_T - n_c) /[ 4g^2 + r(\kappa + \gamma)]^2 $ and in the latter it is $64 \kappa r g^2  (n_c - n_T)/ [(\gamma +\kappa +r)^2 \left(16 g^2 - (\gamma +\kappa -r)^2\right)]$. 

\begin{figure}[tb]
\centering
\includegraphics[width=0.49\textwidth]{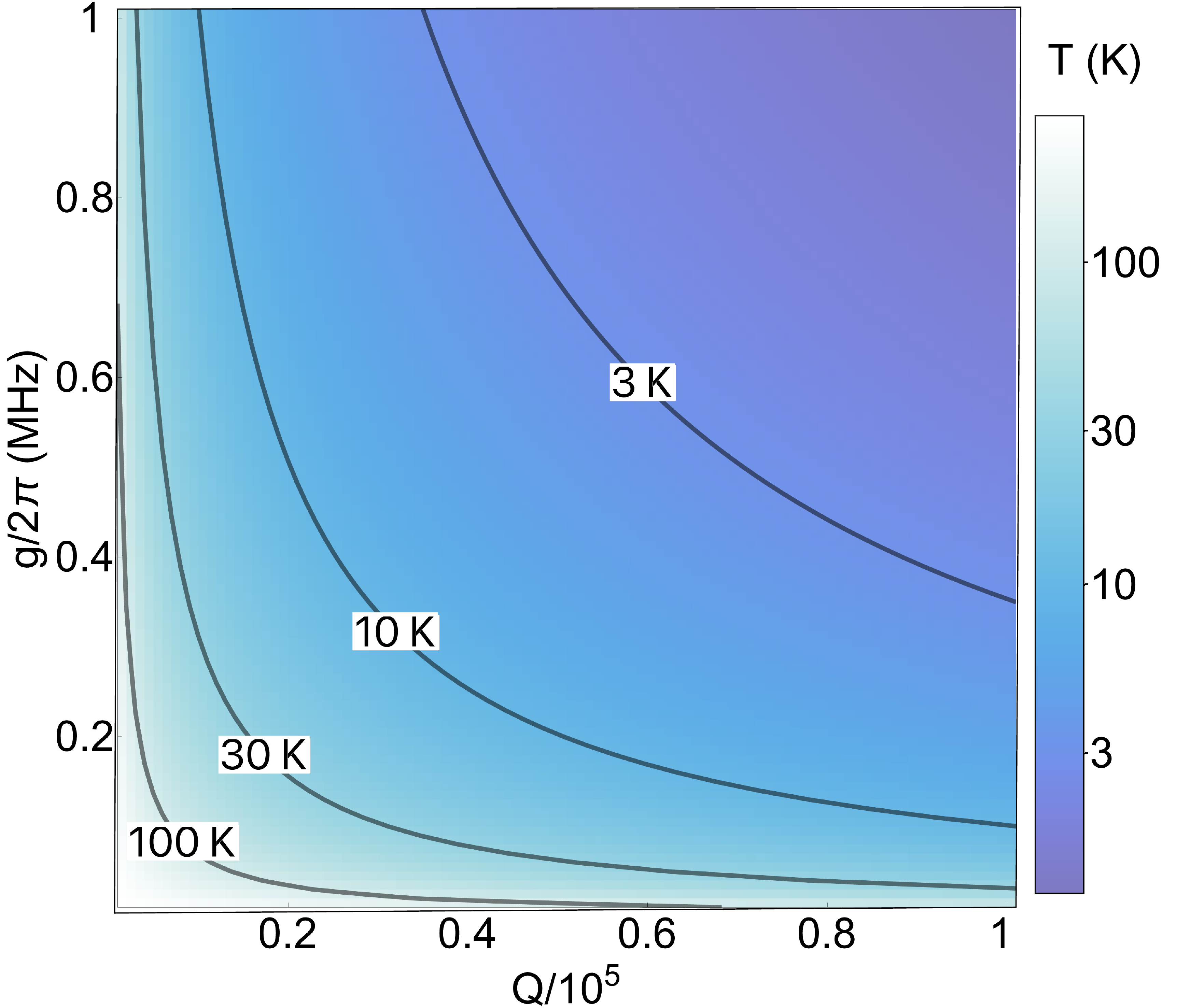}
\caption{Minimum cavity mode temperatures predicted by the linear model (Eq.~\ref{eq:result}) for a range of spin-cavity couplings and cavity $Q = \omega/\gamma$, $\omega = 2.87 \, \textrm{GHz}$ (i.e.\ diamond NV centers), $r = g$, and spin polarization $P = 0.8$, showing that temperatures competitive with cryogenic systems should be achievable with experimentally feasible quality factors and sufficiently strong coupling.}
\label{fig:theory}
\end{figure}

In principle, the ultimate limit for cooling the cavity mode is the temperature of the spin ensemble, which is exceedingly low even for modest spin polarizations (e.g.\ $340 \, \textrm{mK}$ for $P=0.2$ and $\omega = 2 \pi \times 2.87 \, \textrm{GHz})$. Accounting for practical limits on spin-cavity coupling $g$ and cavity quality factor $Q$ (which determines $\gamma$), few-Kelvin mode temperatures are still achievable. To visualize this, we have numerically minimized the noise power in Eq.~\ref{eq:result} over a range of $g$ and $Q$ for the case of $P=0.8$, $r=g$, and cavity mode frequency $\omega = 2 \pi \times 2.87 \, \textrm{GHz}$. The linear model predictions shown in Fig.~\ref{fig:theory} indicate the potential for cryogenic-level cooling with ${\sim} \, \textrm{MHz}$ scale coupling and quality factors of a few tens of thousand.

Our experimental approach used a microwave cavity consisting of two cylindrical stacked dielectric resonators inside a conductive (aluminum) cylindrical shell \cite{Eisenach21}. The resonator halves were made of high dielectric constant, low-loss ceramic sandwiched around a silicon carbide (SiC) substrate supporting the $3\times3\times0.9 \, \textrm{mm}$ CVD-grown diamond sample with 4 ppm total nitrogen-vacancy center concentration \cite{Edmonds2021}.  The TE$_{01\delta}$ mode, chosen for its spatial uniformity and spectral separation from nearby modes, was probed with a shorted loop positioned ${\sim}6 \, \textrm{mm}$ above the resonator. A high power 532 nm pump laser, left on throughout the experiment, continuously polarized the ground state of the diamond NV ensemble spins. Complete details of the setup are in the Methods below and Supplemental Information.

We first characterized the NV-cavity system to estimate the model parameters. With the spins unpolarized (laser off) and not resonant with the cavity, we measured the unloaded cavity mode frequency to be $2891 \, \textrm{MHz}$ with a linewidth of $\gamma = 140 \, \textrm{kHz}$ full width at half maximum (FWHM), corresponding to a quality factor of ${\sim} 2\times 10^{4}$. Introducing the coupling loop and critically coupling to the cavity increased the linewidth by a factor of two. Optically detected magnetic resonance (ODMR) measurements placed an upper bound on the NV spin inhomogeneous linewidth of $290 \,  \textrm{kHz}$.

\begin{figure}[bt]
\centering
\includegraphics[width=0.35\textwidth]{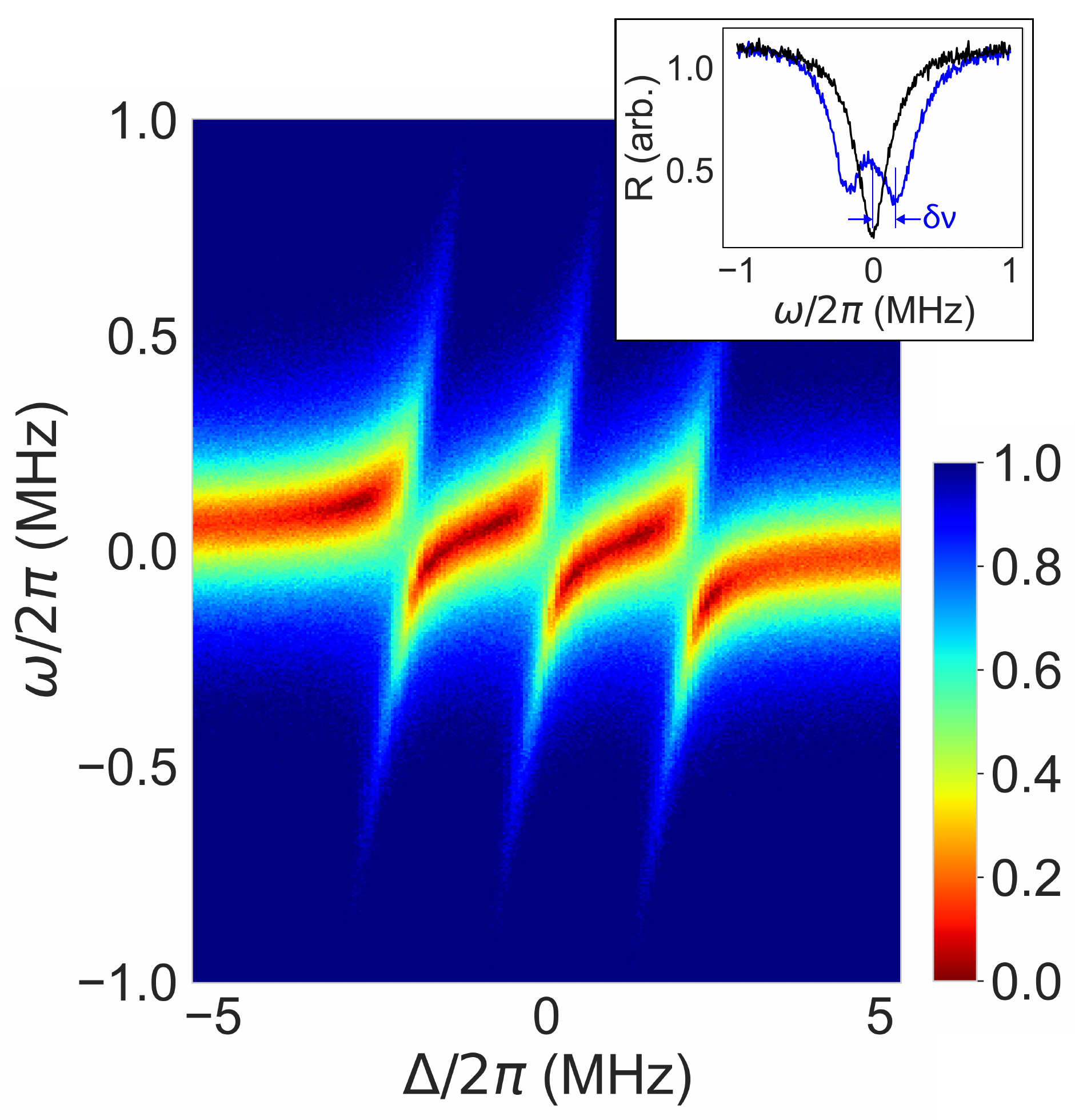}
\caption{The splitting $\delta \nu$ due to the collective coupling of the spin ensemble to the cavity mode was measured by reflecting a weak probe at cavity detuning $\omega$ as the NV spin detuning $\Delta$ was swept through the bare cavity resonance at $2895.3 \, \textrm{MHz}$. The data shown here are for a laser pump power of $5.5 \, \textrm{W}$ distributed over the whole diamond. Three splittings are visible due to the $^{14}$N hyperfine interaction. The inset shows the middle splitting (red dashed line) compared with the bare cavity resonance. }
\label{fig:coupling1}
\end{figure}

Cooling occurs when the NV spins are resonant with the cavity mode. At this resonance, strong coupling between the polarized spin ensemble and the cavity splits the cavity resonance into hybrid spin-cavity eigenmodes. We utilized this cavity splitting to locate the resonance frequency and quantify the amount of coupling for our system. The detuning between the NV spins and the cavity mode ($\Delta$) was controlled by keeping the cavity resonance fixed and scanning the spin transition frequency. Using a uniform magnetic field generated by three sets of orthogonal Helmholtz coil electromagnets aligned to the $\langle 100 \rangle$ crystal axis, all four NV- orientations could be swept through the cavity resonance simultaneously. 

The spin-cavity coupling was measured via the cavity mode splitting. A weak probe signal was reflected off of the cavity and the amplitude spectra (Fig.~\ref{fig:coupling1}) were collected for a range of $\Delta$ by means of microwave homodyne detection. In general, the $^{14}$N-induced hyperfine interaction splits the spin transition into three sub-ensembles separated by roughly $2.15 \, \textrm{MHz}$, resolvable in both our ODMR calibration and cavity reflectivity measurements. The splitting $\delta \nu$ as the magnetic field shifts the spins through the cavity in Fig.~\ref{fig:coupling1} occurs at a distinct bias for each of the three sub-ensembles. Combined with knowledge of the ground state spin polarization, this splitting ($\delta \nu \approx 200 \, \textrm{kHz}$ in Fig.~\ref{fig:coupling1}) quantifies the collective coupling between the low-temperature spin bath and the cavity mode.

\begin{figure*}[t]
\centering
\includegraphics[width=0.95\textwidth]{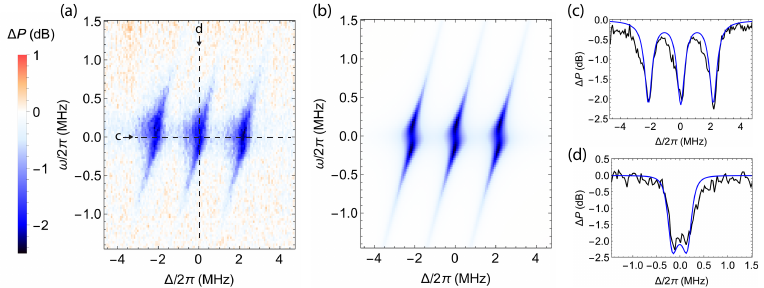}
\caption{Measured change in thermal noise on the cavity output and comparison with the model: (a) steady state change in noise power (compared with no pump) at cavity microwave detuning $\omega$ and NV spin detuning $\Delta$ for an optical pump power of $3 \, \textrm{W}$ and $30 \, \textrm{kHz}$ detection bandwidth, (b) model prediction (Eq.~\ref{eq:result}) assuming a spin polarization $P=0.8$, and parameters $g = 2 \pi \times 197.7 \, \textrm{kHz}$, $\gamma = 2 \pi \times 140 \, \textrm{kHz}$, $r = 2 \pi \times 229.0 \, \textrm{kHz}$, and $\kappa = 2 \pi \times 185.1 \, \textrm{kHz}$. Slices along the NV spin detuning axis and microwave cavity output frequency axis are shown in (c) and (d), respectively, along with the theory prediction from (b). At this pump intensity we report a peak $-2.3 \, \pm 0.1 \, \textrm{dB}$ reduction in thermal noise ($150 \pm 5 \, \textrm{K}$) compared with an unpumped system.}
\label{fig:cooling}
\end{figure*}

Our main experimental result of fully-continuous, spectrally-resolved microwave cavity mode cooling is shown in Fig.~\ref{fig:cooling}. For a cw pump power of $3 \, \textrm{W}$ distributed over the whole diamond and critical output coupling, we report a maximum thermal noise reduction of $-2.3 \pm 0.1 \, \textrm{dB}$, which corresponds to an effective temperature of the cavity output of $150 \pm 5 \, \textrm{K}$ when accounting for detection noise and bandwidth.

The cavity mode cooling measurement was conducted for a range of spin-cavity detunings $\Delta$ but without a probe signal and with the input to the coupling loop circulator terminated. The spin-polarizing pump laser was kept on continuously throughout the experiment. Because the pump laser also heated the diamond through absorption (and thus the resonator through thermal contact), a sufficient amount of time for equilibration (i.e.\ a few minutes) was allowed after any changes to the pump power setpoint. The coupling loop position was then tuned to achieve near-critical coupling to the cavity mode. A three-stub tuner impedance-matched the short (10 cm) coaxial feed from the coupling loop to the detection circuit. This circuit was made up of a low-loss circulator (isolator), two series low-noise high-gain amplifiers ($\textrm{NF}_{\textrm{tot}}<0.8 \, \textrm{dB}$, $G_{\textrm{tot}}\approx +93 \, \textrm{dB}$), and a spectrum analyzer with sufficiently low instrumentation noise. To reduce the effect of instantaneous measurement fluctuations due to phase noise on the spectrum analyzer internal oscillator, 200 independent traces were averaged for each spin detuning.

Thermal noise power goes as $N_{P} = k_{B} T B$ where $B$ represents the measurement bandwidth. In our case this was set by the spectrum analyzer resolution bandwidth which was fixed to $B = 30 \, \textrm{kHz}$, corresponding to $-129.2 \, \textrm{dBm}$ at $290 \, \textrm{K}$. The two-stage, low-noise amplifier further raised this noise floor to $-42.8 \, \textrm{dBm}$, over $100 \, \textrm{dB}$ above the technical floor of the spectrum analyzer, while contributing negligible additional noise. We checked the dominance of thermal noise in our measurement by verifying the bandwidth dependence and agreement with theory (Supplementary Information).

The spectral shape of the mode cooling was discerned through use of a sufficiently narrow detection bandwidth. Figure~\ref{fig:cooling}a clearly shows three well-resolved noise dips as each of the hyperfine sub-ensembles become resonant with the cavity. A fit to our model, shown in Fig.~\ref{fig:cooling}b, accurately reproduces the noise suppression for fixed $\gamma = 2 \pi \times 140 \, \textrm{kHz}$ (as determined from reflectivity measurements) and free parameters $g = 2 \pi \times 197.7 \, \textrm{kHz}$ (similar to the measured cavity splitting), $r = 2 \pi \times 229.0 \, \textrm{kHz}$ (less than but similar to the ODMR broadening), and $\kappa = 2 \pi \times 185.1 \, \textrm{kHz}$ (indicating slight overcoupling). A prefactor accounts for 1.35 dB of loss between the coupling loop and the detector.


The slices shown in Fig.~\ref{fig:cooling}c and Fig.~\ref{fig:cooling}d along the spin and microwave frequency axes, respectively, reveal the sensitivity to tuning mismatch and the cooling bandwidth. From Fig.~\ref{fig:cooling}c it is seen that the noise dips have a $\textrm{FWHM} \approx 670 \, \textrm{kHz}$, roughly 2-3 times the spin ensembles inhomogenous linewidth as measured from ODMR. For a single sub-ensemble tuned to the cavity resonance (Fig.~\ref{fig:cooling}d), the microwave spectrum has a broad, flattened shape with a $\textrm{FWHM} \approx 750 \, \textrm{kHz}$ bandwidth. This is also seen in the model, represented by a solid line and using the same fit parameters as listed above, and indicates the onset of strong coupling. As the spin-cavity coupling is increased, the model predicts a splitting of the cavity mode noise suppression. Our reflectivity measurements suggest that we are on the verge of strong coupling (often designated as when the cavity splitting exceeds the linewidth), and thus we do not resolve a splitting in the cooling but only a broadening.  

\begin{figure}[tb]
\centering
\includegraphics[width=0.5\textwidth]{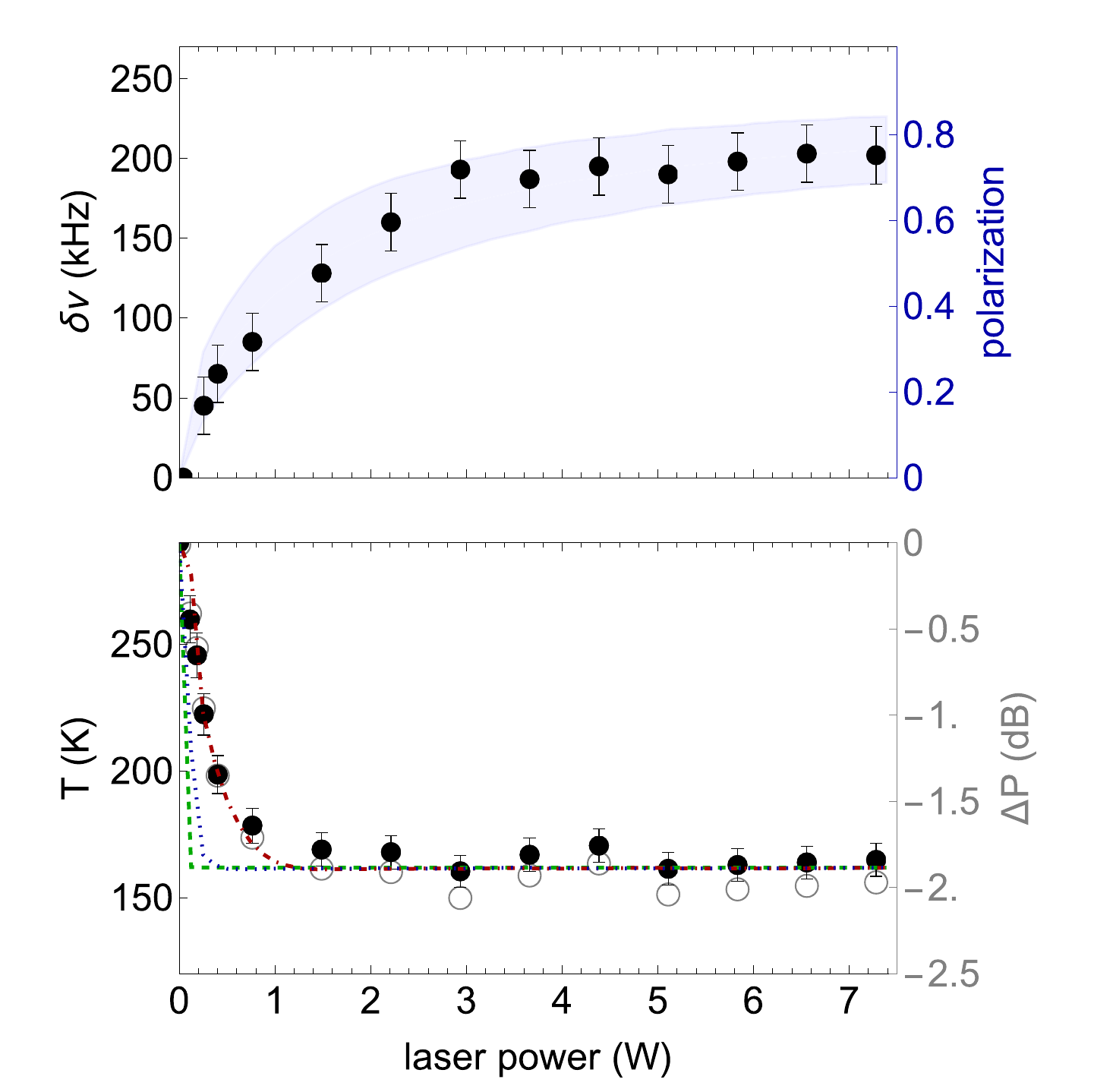}
\caption{Cavity mode splitting $\delta \nu$ (top) and depth of cavity mode cooling (bottom) versus applied optical pump power. The mode splitting, itself a measure of spin-cavity coupling $g$, depends partly on the spin polarization generated by the pump laser. The predicted behavior from a model including temperature effects is shown by the blue region (Supplementary Information). The bottom plot shows the inferred cavity mode temperature from measuring the noise power reduction on the output (black circles). Error bars indicate standard error. The green dashed line shows the prediction by the linearized model, and the blue dot-dashed line includes a phenomenological correction to the spin-cavity coupling.}
\label{fig:rate}
\end{figure}

The effective cooling rate depends critically on the spin-cavity coupling $g$ (Eq.~\ref{eq:result}) as well as the spin polarization $P$ (Eq.~\ref{eq:result}) which sets the effective temperature of the NV spin ensemble. Although the bare $g$, which is partly determined by NV${-}$ center concentration and cavity mode overlap with the diamond, is not easily changed in our experimental system, we could control the optical pump intensity with a variable attenuator which had a direct impact on $P$. We varied the pump laser intensity over two orders of magnitude and measured its effect on the cavity mode splitting and mode temperature. The results (Fig.~\ref{fig:rate}) display an increase in both cavity splitting and mode cooling with higher laser power followed by a leveling off at about $3\, \textrm{W}$ or $10^{-4} \, \textrm{mW}/\mu\textrm{m}^2$ optical intensity. 

We explain these phenomena through a two-level NV rate model \cite{PhysRevB.84.195204} including the competing effects of optical spin pumping and temperature-dependent $T_1$ relaxation. The blue filled region in Fig.~\ref{fig:rate} indicates the results of a Monte Carlo simulation for the spin polarization using the above model with uncertainties from the literature and estimated for our sample. The correlation seen in Fig.~\ref{fig:rate} between the cavity mode splitting (i.e.\ spin-cavity coupling) and ground state polarization is explained by the former depending on the number of emitters coherently interacting with the cavity mode. As the optically-induced spin pumping rate of the NV ensemble becomes comparable to the longitudinal relaxation rate (1/$T_1$), the polarization begins to saturate (Fig.~\ref{fig:rate} top). As the excitation laser beam was spread over a diameter of ${\sim}$5 mm, leading to pumping intensities ($<10^{-4} \, \textrm{mW} /  \upmu \textrm{m}^{2}$) much smaller than the NV saturation intensity (${\sim} 1 \, \textrm{mW} /  \upmu \textrm{m}$) \cite{PhysRevB.84.195204}, non-linear processes such as photoionization and charge state conversion are negligible. The steady state polarization reached here was lower than that achieved in single-NV experiments \cite{PhysRevB.84.195204} (Supplementary Information), but as spin polarization is exponentially related to spin temperature, the limited polarization does not preclude microwave mode cooling. The dashed green line shows the model prediction (Eq.~\ref{eq:result}) with the fitted parameters from Fig.~\ref{fig:cooling} and matches the data for high polarization but fails to capture the data at small polarizations. The blue dashed line shows a phenomenological model where the effective $g$ varies with the square-root of the $P$, while the best agreement comes from assuming $g \propto P$ shown as the red dot-dashed line. Further theoretical work is required to understand this behavior at low optical pump powers.

\begin{figure}[tb]
\centering
\includegraphics[width=0.4\textwidth]{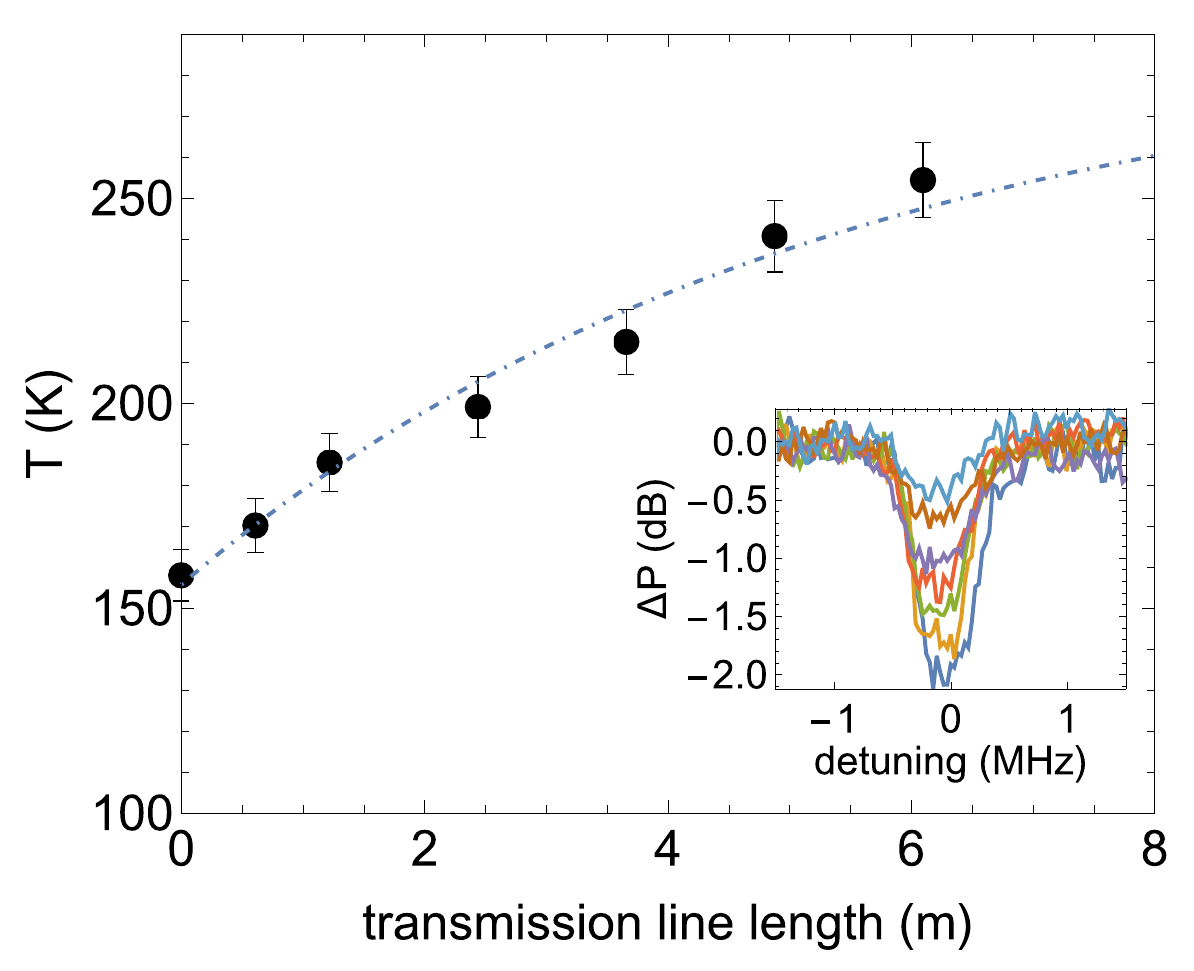}
\caption{Varying the length of coaxial cable between the output of the cavity and the detector. We find that the rate of cooling decreases exponentially with a length constant of $\alpha = 5.3 \, \textrm{m}$. Inset: Spectrally-resolved power reduction at varied cable lengths.}
\label{fig:cable}
\end{figure}

The microwave mode cooling produced in the cavity can be transferred to a transmission line, producing non-thermal voltage power spectral density at a distance. A demonstration is shown in Fig.~\ref{fig:cable} where we vary the length of a coaxial microwave line from 0 to 6 m and measure the effective temperature from the noise power as before. We find that the cooling in the cavity band falls exponentially with cable length, with constant $\alpha = 5.3 \, \textrm{m}$ consistent with the ohmic loss in the line ~\cite{Dimitrios87}. At a distance of 6 m we observe a reduction of 40 K, showing the potential of this technique for cooling on laboratory scales.

The results presented here open a pathway to reducing noise in a range of applications as well as to achieving much lower temperatures. The latter could enable entirely new applications, for example in quantum technology at ambient conditions. There are a number of routes to realising lower temperatures. The number of spins could be increased by using a larger diamond sample or one with greater NV- concentration. The effective spin/cavity coupling could be increased using new cavity designs that concentrate the mode at the location of the ensemble while maintaining high $Q$~\cite{choi2021ultrastrong}. Importantly, as our theoretical model reveals, cooling requires sufficient inhomogeneous broadening of the spins. Assuming one starts with an underbroadened NV ensemble, tuning this inhomogeneous linedwidth, for example with a magnetic field gradient, will allow one to optimize the thermal noise suppression. As it stands the continuous cooling demonstrated in this work surpasses what is possible with thermoelectrically-cooled systems, and offers strong potential for performance competitive with cryogenic systems in the future.

\section{Methods}
The purple diamond ($3\times 3\times 0.9 \, \textrm{mm}$, $\textrm{NV} >4 \, \textrm{ppm}$) was situated on a $330 \, \upmu \textrm{m}$ thick, two-inch diameter SiC wafer (4H semi-insulating, MSE Supplies) which provided a thermally conductive path to the shield. Pump laser light came from a diode-pumped solid-state (DPSS) laser providing up to $10 \, \textrm{W}$ of continuous-wave power at $532 \, \textrm{nm}$ (Lighthouse Photonics Sprout-G), and its intensity was controlled with a motorized laser power attenuator (Optogama LPA). The beam was expanded to a $1/e^{2}$ diameter of $5 \, \textrm{mm}$ to improve the pump homogeneity across the sample, and apertured to a $6.5 \, \textrm{mm}$ diameter to avoid unwanted heating from beam edges. A dichroic mirror was used as the last reflector before the microwave shield to allow for the collection of photoluminescence for ODMR and microwave-assisted charge state spectroscopy \cite{AudeCraik2020}. This in-situ ODMR was used for the diamond cavity system to calibrate between NV spin shift values for a given Helmholtz coil bias current.

The cavity mode was probed with a shorted 6 mm diameter loop formed from the center conductor of one end of a 10 cm piece of coaxial cable (Pasternack RG405, $0.5 \, \textrm{dB}$ loss per foot), and concentrically aligned with the dielectric resonator 6 mm from one end. The loop feed was connected to a three-stub tuner (Maury Microwave) and a low-loss circulator (RF-Lambda RFLC313G27G37, 0.6 dB loss) by which optional probe signals could be reflected off of the cavity or the cavity mode noise could be read out. The $\textrm{TE}_{01 \delta}$ mode was characterized by monitoring the S21 parameter through the circulator using a Keysight N9915A microwave analyzer in network analyzer mode. The degree of loop-cavity coupling could be continuously adjusted via the loop position (e.g. translation stage) and stub tuner, with the reflected Lorentzian linewidth FWHM asymptotically approaching 140 kHz ($Q \approx 2 \times 10^4$) and measuring ${\sim}230$ kHz at critical coupling. Fine adjustments to the loop position were needed to recover critical coupling when the optically pumped NV ensemble was resonant with the cavity. To verify the cavity measurements, we simulated the system with finite element methods using COMSOL multiphysics electromagnetic wave module. The mode frequency and linewidth from these simulations were in excellent agreement with measurements, and further details are given in the Supplementary Information.

The cavity mode noise was measured with a Keysight N9915A microwave analyzer in spectrum analyzer mode. The adjustable resolution bandwidth $B$ set the frequency range over which the flat thermal noise was integrated, corresponding to a theoretical thermal noise floor of $N_{P} = k_{B} T B$ with Boltzmann constant $k_{B}$ and temperature $T$. For $B = 30 \, \textrm{kHz}$ at $300 \, \textrm{K}$, this corresponds to a power level of $-129 \, \textrm{dBm}$. The noise figure of the spectrum analyzer of $22.5 \, \textrm{dB}$ set the actual floor to $-106.5 \, \textrm{dBm}$. To operate well-above instrumentation noise we used two low-noise amplifiers in series: the first with $55.7 \, \textrm{dB}$ gain and noise figure $0.75 \, \textrm{dB}$ (RF-Lambda RLNA02G04G60) and the second with $37.1 \, \textrm{dB}$ gain and noise figure $0.8 \, \textrm{dB}$ (RF-Lambda RLNA02G08G30). Together these elevated the thermal noise of the cavity mode to $-42.8 \, \textrm{dBm}$. Precise determination of the noise floor was accomplished by averaging a few hundred traces to eliminate instantaneous phase fluctuations on the internal oscillator of the spectrum analyzer.

\begin{acknowledgments}
The authors would like to thank Danielle Braje and Erik Eisenach for helpful discussions. M.E.T. acknowledges support from the ARL ENIAC Distinguished Postdoctoral Fellowship. H.C. and D.E acknowledge the support from DARPA DRINQS. M.J.T acknowledges support from the Army Research Laboratory Maryland-ARL Quantum Partnership program under Contract No. W911NF-19-2-0181 and the University of Maryland Quantum Technology Center.
\end{acknowledgments}

\end{document}